\documentclass[doublespacing]{elsart}
\usepackage{stmaryrd}
\usepackage{amsmath}
\usepackage{amssymb}
\usepackage{amsfonts}
\usepackage{graphicx}
\usepackage{ctex}
\usepackage{rotating}
\usepackage{longtable}
\date{}
\begin{document}

\noindent \textbf{Corresponding author: }\\
 \noindent Dr. Hongjian Feng\\
\noindent School of Science,Tianjin Polytechnic University, Tianjin 300160, China\\
\noindent Department of Physics and Astronomy, University of
Missouri,Columbia,  MO 65211, USA\\
\noindent Email address:\\
fenghongjian@gmail.com\\
fenghongjian1978@gmail.com
 \clearpage
\begin{frontmatter}



\title{The role of Coulomb and exchange interaction on the Dzyaloshinskii-Moriya interaction(DMI) in BiFeO$_3$}

\author[label1,label2]{Hongjian  Feng}
 \address[label1]{School of Science,Tianjin Polytechnic University, Tianjin 300160, China}
\address[label2]{Department of Physics and Astronomy, University of
Missouri,Columbia,  MO 65211, USA} \ead{fenghongjian@gmail.com}


\begin{abstract}
$Ab$ $initio$ calculations show that the Dzyaloshinskii-Moriya
interaction(DMI)and net magnetization per unit cell in BiFeO$_3$ are
reduced when  $U$ is increasing from 0 to 2.9 eV, and independent of
$J$. Interestingly, the DMI is even destroyed as $U$ exceeds a
critical value of 2.9 eV. We propose a simple model to explain this
phenomenon and present the nature of the rotation of the
magnetization  corresponding to altered antiferrodistortive
distortions under DMI in BiFeO$_3$.

\end{abstract}

\end{frontmatter}


\section{ Introduction}
Multiferroic materials have attracted much interest due to the
coexistence of magnetic and ferroelectric ordering in single phase.
The coupling of the two ordering leads to the so-called
magnetoelectric effect in which the magnetization can be tuned by
the external electric field, and vice versa\cite{1,2,3,4,5}. These
materials have potential applications in information storage, the
emerging field of spintronics,and sensors. BiFeO$_3$  is the rare
one in nature, which possess both weak ferromagnetism and
ferroelectric characteristics in single phase\cite{6,7,8,9}. It has
long been known to be ferroelectric with a Curie temperature of
about 1103 K and antiferromagnetic(AFM) with a N\'{e}el temperature
of 643 K. The Fe magnetic moments are coupled ferromagnetically in
(1 1 1) plane and antiferromagnetically in the adjacent plane along
[111] direction, which is known as the G-type AFM order.
 The rhombohedral distorted perovskite
structure with space group $R3c$ permits a canting of AFM sublattice
caused by the antisymmetric Dzyaloshinskii-Moriya interaction(DMI),
resulting in a weak ferromagnetism. However, there is a spiral spin
structure in which the AFM axis rotates through the crystal with a
 long-wavelength period of 620{\AA}.
The cancellation of magnetization should be suppressed partly in
thin film \cite{6}, or by partly substitution of  magnetic
transitional metal ions in  B sites, as shown in our previous
report\cite{10}. It is known that the ferroelectricity in BiFeO$_3$
is produced by the lone Bi-6s stereochemically  active pair induced
by the mixing between the $(ns)^2$ ground state and  a low-lying
$(ns)^1(np)^1$ excited state, which can only occur if the cation
ionic site does not have inversion symmetry, while the weak
ferromagnetism is mainly attributed to Fe$^{3+}$ ions. Therefore the
coupling between the electric  and the magnetic ordering becomes
weak in BiFeO$_3$, which agrees with the fact of large difference
between the Curie temperature  and AFM N\'{e}el temperature. There
exists another structural distortion, so-called
antiferrodistortive(AFD) distortion, which is formed by the
alternating sense of rotation of the oxygen octahedra along [1 1 1]
direction\cite{11}. In our previous paper, we have shown that the
rotation of the oxygen octahedra couples with the weak
ferromagnetism due to the DMI, using $Ab$ $initio$ calculations with
considering the spin-orbital(SO) coupling effect and the
noncollinear spin configuration\cite{12}. In strongly correlated
materials, e.g. multiferroics, the on-site Coulomb($U$) and exchange
interaction($J$) has been proposed to properly describe the partly
filled localized $d$ orbitals within density functional theory(DFT).
One may wonder whether $U$ and $J$ will have an impact  on the
magnetization through DMI taking into account SO interaction. How do
these parameters influence the DMI, and further the magnetization?
What is the origin of  coupling between the rotation of oxygen
octahedra and the resulting magnetization in terms of DMI  in
BiFeO$_3$? In this paper we have proposed a transparent physical
interpretation for the abovementioned questions,using
first-principles calculations based on the DFT.

The remainder of this article is structured as follows:In section 2,
we presented the computational details of our calculations.  We
provided the calculated results and discussions in section 3. In
section4, the conclusion based on our calculation were given.

\section{ Computational details}

 Our calculations were performed within the local spin density
 approximation(LSDA) to DFT using the
ABINIT package\cite{13,14}.  The ion-electron interaction was
modeled by the projector augmented wave (PAW) potentials
\cite{15,16} with a uniform energy cutoff of 500 eV. Bi 5d, 6s, and
6p electrons, Fe 4s, 4p,and 3d electrons, and O 2s and 2p electrons
were considered as valence states. Two partial waves per $l$ quantum
number were used. The cutoff radii for the partial waves for Bi, Fe,
and O were 2.5, 2.3, 1.1 a.u., respectively. $6\times6\times6$
Monkhorst-Pack sampling of the Brillouin zone were used for all
calculations. We calculated the net magnetization per unit cell and
the electronic properties within the LSDA+U method where the strong
coulomb repulsion between localized $d$ states has been considered
by adding a Hubbard-like term to the effective
potential\cite{17,18,19}. The effective Hubbard parameter, the
difference between  the Hubbard parameter $U$ and the exchange
interaction $J$ ($U-J$), was changing in the range between 0 and 6
eV for the Fe $d$ states. For the same value of $(U-J)$, $J$ was
varying as 0,0.5, 0.8,and 1 eV, respectively. Taking into account
the SO interaction, we introduced the noncollinear spin
configuration to construct the G-type AFM magnetic order with the
AFM axis being along the $x$ axis in Cartesian coordinates  in our
$Ab$ $initio$ calculation.

\section{ Results and discussion}

In ref. 20 the author suggest that the inversion centers between
adjacent B sites in ABO$_3$ perovskite structure are destroyed by
the displacement of the oxygen anions located at the midpoints
between them, while the space inversion centers between A sites
still remains. Therefore ABO$_3$ structure with magnetic ions in A
sites, such as FeTiO$_3$, should possess a strong coupling between
the ferroelectric distortions and magnetization. It can not be
achieved in ABO$_3$ structure with magnetic ions in B sites, such as
BiFeO$_3$. That is to say the coupling between the ferroelectric
distortions and magnetization in it shall be neglected.  However in
BiFeO$_3$ there exists another kind displacement, known as
antiferrodistortive(AFD) distortions, caused by the rotation of the
neighboring oxygen octahedra. Through antisymmetric superexchange
interaction this AFD displacement couples weakly to the
magnetization. In this paper we mainly concentrate on the coupling
associated with the DMI between the AFD distortions and the
magnetization per unit cell.

For the AFD motion, a rotational vector $\mathbf{R}$ has been
introduced to describe the direction of the rotation of the oxygen
octahedra\cite{12}. From Fig. 1, the anticlockwise rotation of upper
oxygen octahedra and clockwise rotation of lower oxygen octahedra
correspond to the outward state defined as $\mathbf{R}_{out}$. The
opposite state is defined as $\mathbf{R}_{in}$. The rotational angle
is 10\textordmasculine   in the Cartesian coordinates\cite{12}.

In our LSDA+U calculation, $U$ and $J$ are defined as
\begin{equation}
U=\frac{1}{(2l+1)^2}\sum_{m,m'}<m,m'|V_{ee}|m,m'>=F^0,
\end{equation}
\begin{equation}
J=\frac{1}{2l(2l+1)}\sum_{m\neq
m',m'}<m,m'|V_{ee}|m,m'>=\frac{F^2+F^4}{14},
\end{equation}
where $V_{ee}$ are the screened Coulomb interaction among the $nl$
electrons. $F^0$, $F^2$, and$F^4$ are the radial Slater integrals
for $d$ electrons in Fe.

The net magnetization per unit cell with respect to
$\mathbf{R}_{in}$ and $\mathbf{R}_{out}$ in Cartesian coordinates
for different $U$ and $J$ were listed in table 1. It can be seen
that $J$ value have no effect on the resulting magnetization when
$U$ remains constant. For the sake of clarity, only the results
obtained with different $J$ value for $U$=0 and 2.9 eV were given in
the table . The AFM vector in Cartesian coordinates with varying
effective Hubbard $U$ were illustrated in Fig. 2, where [1 1 1]
direction is taken as the $z$ axis as shown schematically in Fig. 3,
and the $x,y,$and $z$ component of the magnetization is denoted by
$M_x$, $M_y$, and $M_z$ in the coordinates, respectively. Fig. 3
shows the coupling between the rotation of oxygen octahedra and the
resulting magnetization per unit cell. The arrow indicate the spin
direction of Fe for different states. The upper section corresponds
to the R$_{in}$ rotational state, and lower section, the R$_{out}$
rotational state. M$_{total}$ is the net magnetization per unit
cell. The dashed line arrow is the unstable rotational state. [1 1
1] crystal direction is selected as the $z$ axis, and the AFM order
is arranged along x axis.It is clearly shown that the easy axis of
the magnetization is $y$ axis when G-type AFM order is arranged
along the $x$ axis, taking into account the SO interaction and the
unconstrained freedom of spin. The antisymmetric interaction of the
neighboring Fe1 and Fe2 ions leads to the canting of the magnetic
moment of them away from their original direction($x$ axis) and a
resulting magnetization mainly in $y$ axis, which arises from the
DMI only occurring when the inversion symmetry is broken. As U is
approaching from 0 eV to 2.9 eV,the net magnetization is reversed by
the opposite rotation of the oxygen octahedra in terms of the
reversal of $M_y$, and decreases with increasing of $U$. However,
$M_y$ does not change sign with the altered AFD motion when $U$
exceeds a critical value of 2.9 eV, say 3 eV, implying that the net
magnetization only deviates slightly from the original direction and
does not experience a significant rotational angle greater than
90\textordmasculine.As U attain to be the critical value, the DMI
caused by the antisymmetric superexchange interaction is eliminated
with the strong on-site Coulomb interaction. The AFD distortions do
not couple with the magnetization.

 In order to obtain an unambiguous interpretation for
the effect of Coulomb and exchange interaction on the net
magnetization, we need to recap the DMI on the coupling of
neighboring Fe1 and Fe2 sites. We have for the interaction of
neighboring Fe1 and Fe2 sites by the second order perturbation
calculation\cite{21,22,23}

\begin{equation}
E_{Fe1,Fe2}^{(2)}=\mathbf{J}_{Fe1,Fe2}^{(2)}(\mathbf{S_1}\cdot\mathbf{S_2})+
\mathbf{D}_{Fe1,Fe2}^{(2)}(\mathbf{S_1}\times\mathbf{S_2})+\mathbf{S}(R)\cdot
\Gamma_{Fe1,Fe2}^{(2)}\cdot \mathbf{S}_2.
\end{equation}
The first term on the right hand side of the Eq. (3) corresponds to
the usual isotropic superexchange interaction, and the second term
is the DMI.  Provided the long range pseudodipolar interaction is
neglected, we get the Hamiltonian for the system

\begin{equation}
H_{BiFeO_3}=-2\sum_{<1i,2j>}\textbf{J}_{1i,2j}\mathbf{S}_{1i}\cdot\mathbf{S}_{2j}+\sum_{<1i,2j>}\textbf{D}_{1i,2j}\mathbf{S}_{1i}\times\mathbf{S}_{2j}.
\end{equation}
The first term comes from the symmetric superexchange, and the
second one is the antisymmetric DMI contribution.
$\textbf{J}_{1i,2j}$ in the first term is a constant similar to the
exchange interaction, and does not contribute to the DMI. This can
account well for our calculated results  that the exchange parameter
$J$ has nearly no effect on the rotation of the magnetization.
\textbf{D} is the DMI constant associated with the crystal field and
determined by the sense of rotation of the neighboring oxygen
octahedra(\textbf{R}$_{in}$ or \textbf{R}$_{out}$). \textbf{D} reads
by the second order perturbation in the case of one electron per ion
\begin{equation}
\textbf{D}_{Fe1,Fe2}^{(2)}=(4i/U)[b_{nn'}(Fe1-Fe2)C_{n'n}(Fe2-Fe1)-C_{nn'}(Fe1-Fe2)b_{n'n}(Fe2-Fe1)],
\end{equation}
where $U$ is the energy required to transfer one electron from one
site to its nearest neighbor, a parameter similar to on-site Coulomb
interaction in our $Ab$ $initio$ computation, and inversely
proportional to \textbf{D}. This is consistent with our calculated
results that the absolute value of net magnetization is inversely
proportional to the Hubbard parameter $U$. Magnetization does not
reverse its direction in terms of the changing of the AFD
displacement, especially when $U$ is greater than the critical value
of 2.9 eV , this indicates that in this case $U$ is large enough to
make DMI being disappeared. We have also calculated the band gap for
different $U$ corresponding to \textbf{R}$_{in}$ and
\textbf{R}$_{out}$, respectively. From Fig. 4, it can be seen that
the curve becomes relatively flat when $U$ reaches the critical
value of 2.9 eV. Thereafter, we chose this value to describe the
electronic property in the following. It is worth mentioning that
the band gap to \textbf{R}$_{in}$ is greater than to
\textbf{R}$_{out}$, indicating that the AFD motion corresponding to
\textbf{R}$_{out}$ tend to reduce the crystal-field splitting , and
consequently the band gap.

In order to analyze the rotation of magnetization under DMI, we have
calculated the Orbital-resolved density of states(ODOS) for Fe1 and
Fe2 corresponding to \textbf{R}$_{in}$ and \textbf{R}$_{out}$ in
Fig.5, Fig. 6, Fig. 7, and Fig 8, respectively. Fig. 5 is the ODOS
for Fe1 corresponding to \textbf{R}$_{in}$ rotational state. The
vertical line indicates the Fermi level. All the states occupied in
the valence band are spin-up electrons(majority spin as defined). It
means the spin direction for Fe1 to \textbf{R}$_{in}$ is positive.
Fig. 6 is the ODOS for Fe2 corresponding to \textbf{R}$_{in}$
rotational state. The vertical line indicates the Fermi level. All
the states occupied in the valence band are spin-down
electrons(majority spin as defined). It means the spin direction for
Fe2 to \textbf{R}$_{in}$ is negative. Fig. 7 is the ODOS for Fe1
corresponding to \textbf{R}$_{out}$ rotational state. The vertical
line indicates the Fermi level. All the states occupied in the
valence band are spin-up electrons(majority spin as defined). It
means the spin direction for Fe1 to \textbf{R}$_{out}$ is positive.
Fig. 8 is the ODOS for Fe2 corresponding to \textbf{R}$_{out}$
rotational state. The vertical line indicates the Fermi level. All
the states occupied in the valence band are spin-down
electrons(majority spin as defined). It means the spin direction for
Fe2 to \textbf{R}$_{out}$ is negative. Let us come back to Fig.3. In
order to make the net magnetization reversed, the magnetic moment of
Fe1 and Fe2 can rotate from the original direction corresponding to
\textbf{R}$_{in}$(Fig. 3(a)) either to the dashed line arrow(Fig.3
(b)) required greater energy barrier, or to the real line arrow
required smaller energy barrier. From Fig. 5 to Fig.8, one can see
that the  spin-up electrons in the occupied valence band for Fe1 and
the spin-down electrons in the occupied valence band for Fe2 do not
change their in-built spin direction when rotational vector is
changing   from \textbf{R}$_{in}$ to \textbf{R}$_{out}$. This
confirms that the spin direction of Fe1 and Fe2 only deviates
slightly from the initial states to the final states as shown in the
real line arrow in Fig.3 (b) corresponding to \textbf{R}$_{out}$. It
is worth pointing out that $d_{x^2-y^2}$ orbital for Fe1 and Fe2 is
split from the doubly degenerate $e_g$ states and tend to overlap
with $d_{xy},d_{yz},$ and $d_{xz}$ orbitals in the triply degenerate
$t_{2g}$ states, indicating that the AFM DMI is mainly attributed to
the $e_g$-$e_g$ AFM interaction which is greater than the
$t_{2g}$-$t_{2g}$ AFM interaction.

\section{ Conclusion}
Magnetization can be reversed by the altering sense of rotation of
the oxygen octahedra in BiFeO$_3$ when $U$ is smaller than the
critical value of 2.9 eV, and the absolute value of magnetization is
decreasing as $U$ is ranging from 0 to 2.9 eV. Magnetization does
not reverse with altered AFD displacement when $U$ exceeds the
critical value, indicating that the DMI is even prohibited in this
case. The rotation of magnetization is fulfilled by slight deviation
of magnetic moment of Fe1 and Fe2 around $x$ axis rather than
reversal of them.



\clearpage

\begin{table}[!h]

 \caption{ Magnetization per unit cell with respect to different value of $U$ and$J$.}

\begin{center}

\begin{tabular}{@{}ccccccccccc}
\hline\hline
$U(eV)$&\multicolumn{2}{c}{0}&\multicolumn{2}{c}{0.5}&\multicolumn{2}{c}{0.8}&
\multicolumn{2}{c}{1}&\multicolumn{2}{c}{1}\\
$J(eV)$&\multicolumn{2}{c}{0}&\multicolumn{2}{c}{0.5}&\multicolumn{2}{c}{0.8}&
\multicolumn{2}{c}{1} &\multicolumn{2}{c}{0}\\
&$\mathbf{R_{in}}$&$\mathbf{R_{out}}$&$\mathbf{R_{in}}$&$\mathbf{R_{out}}$&$\mathbf{R_{in}}$&
$\mathbf{R_{out}}$&$\mathbf{R_{in}}$&$\mathbf{R_{out}}$&$\mathbf{R_{in}}$&$\mathbf{R_{out}}$\\
\hline
$M_x(\mu_B)$&0.0000&0.0000&0.0000&0.0000&0.0000&0.0000&0.0000&0.0000&0.0000&0.0000\\
$M_y(\mu_B)$&0.4259&-0.0812&0.4259&-0.0812&0.4259 &-0.0812&0.4259&-0.0812&0.0351 &-0.0679\\
$M_z(\mu_B)$&-0.1013&0.0000&-0.1013&0.0000&-0.1013&0.0000&-0.1013&0.0000&-0.0493&0.0056\\
\hline\hline
$U(eV)$&\multicolumn{2}{c}{2}&\multicolumn{2}{c}{2.5}&\multicolumn{2}{c}{2.8}&
\multicolumn{2}{c}{2.9}&\multicolumn{2}{c}{3.4}\\
$J(eV)$&\multicolumn{2}{c}{0}&\multicolumn{2}{c}{0}&\multicolumn{2}{c}{0}&\multicolumn{2}{c}{0}&
\multicolumn{2}{c}{0.5}\\
&$\mathbf{R_{in}}$&$\mathbf{R_{out}}$&$\mathbf{R_{in}}$&$\mathbf{R_{out}}$&$\mathbf{R_{in}}$&
$\mathbf{R_{out}}$&$\mathbf{R_{in}}$&$\mathbf{R_{out}}$&$\mathbf{R_{in}}$&$\mathbf{R_{out}}$\\
\hline
$M_x(\mu_B)$&0.0000&0.0000&0.0000&0.0000&0.0000&0.0000&0.0000&0.0000&0.0000&0.0000\\
$M_y(\mu_B)$&0.0416&-0.0337&0.0325&-0.0251&0.0365&-0.0188&0.0313&-0.0176&0.0313&-0.0176\\
$M_z(\mu_B)$&-0.0408&0.0108&-0.0366&0.0147&-0.0431&0.017&-0.0406&0.0168&-0.0406&0.0168\\
\hline\hline $U(eV)$&\multicolumn{2}{c}{3.7}&\multicolumn{2}{c}{3}&
\multicolumn{2}{c}{4}&\multicolumn{2}{c}{5}& \multicolumn{2}{c}{6}\\
$J(eV)$&\multicolumn{2}{c}{0.8}&\multicolumn{2}{c}{0}&
\multicolumn{2}{c}{0}&\multicolumn{2}{c}{0}&\multicolumn{2}{c}{0}\\
&$\mathbf{R_{in}}$&$\mathbf{R_{out}}$&$\mathbf{R_{in}}$&$\mathbf{R_{out}}$&
$\mathbf{R_{in}}$&$\mathbf{R_{out}}$&$\mathbf{R_{in}}$&$\mathbf{R_{out}}$&$\mathbf{R_{in}}$&$
\mathbf{R_{out}}$\\
\hline
$M_x(\mu_B)$&0.0000 &0.0000  &0.0000 &0.0000&0.0000 &0.0000  &0.0000 &0.0000&0.0000 &0.0000\\
$M_y(\mu_B)$&0.0313&-0.0176&0.0237&0.0012&0.0172&0.0033& 0.0176&0.0085 &0.0178 & 0.0111\\
$M_z(\mu_B)$&-0.0406&0.0168&-0.0283&-0.0049&-0.0249&-0.0055&-0.0186& -0.0053  &-0.0157 &-0.0025\\
 \hline\hline
\end{tabular}

\end{center}

\end{table}

\clearpage

\raggedright \textbf{Figure captions:}

Fig.1 The rotational vectors of the AFD distortions in BiFeO$_3$.
The shaded cage denote the oxygen octahedra, and Fe is inside the
cage.

 Fig.2 AFM vectors with respect to $U$.

 Fig.3 Schematic diagram for the coupling between the rotation of
 oxygen octahedra and the resulting magnetization in unit cell in
 BiFeO$_3$. The arrow denote the direction of magnetization.

 Fig. 4 Band gap for \textbf{R}$_{in}$ and
\textbf{R}$_{out}$ with respect to $U$.

 Fig. 5 ODOS for Fe1
$d_{xy},d_{yz},d_{z^2},d_{xz}$, and $d_{x^2-y^2}$ orbitals to
\textbf{R}$_{in}$.

 Fig. 6 ODOS for Fe2
$d_{xy},d_{yz},d_{z^2},d_{xz}$, and $d_{x^2-y^2}$ orbitals to
\textbf{R}$_{in}$.

 Fig. 7 ODOS for Fe1
$d_{xy},d_{yz},d_{z^2},d_{xz}$, and $d_{x^2-y^2}$ orbitals to
\textbf{R}$_{out}$.

 Fig. 8 ODOS for Fe2 $d_{xy},d_{yz},d_{z^2},d_{xz}$, and $d_{x^2-y^2}$
orbitals to \textbf{R}$_{out}$.

\clearpage
\begin{figure}
\includegraphics[width=8cm]{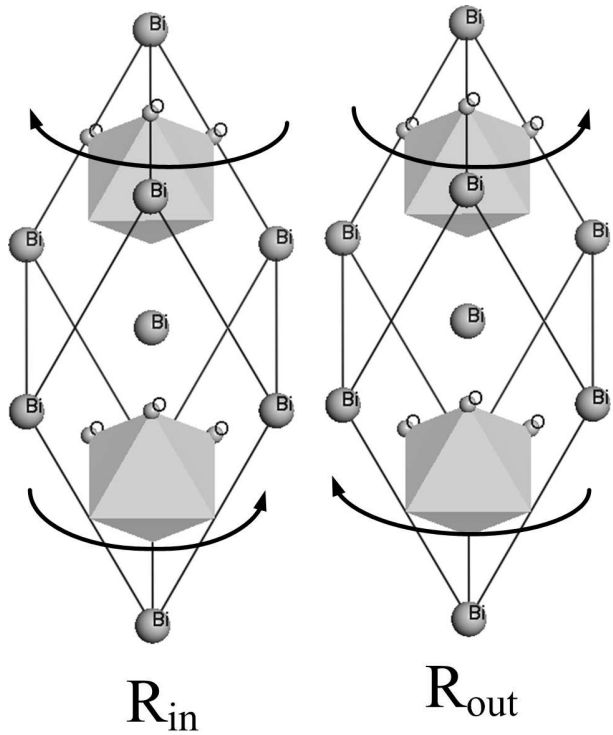}
\caption{The rotational vectors of the AFD distortions in
BiFeO$_3$.The shaded cage denote the oxygen octahedra, and Fe is
inside the cage.}
\end{figure}

\begin{figure}
\includegraphics{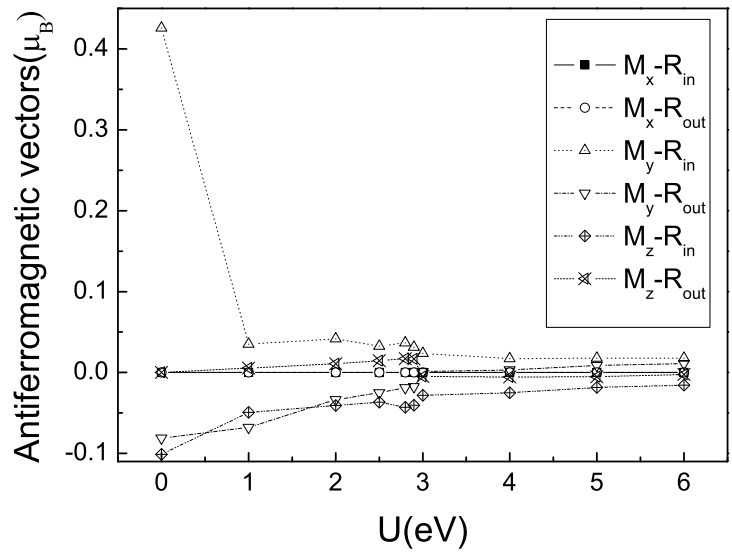}
\caption{ AFM vectors with respect to $U$.}
\end{figure}

\begin{figure}
\includegraphics[width=8cm]{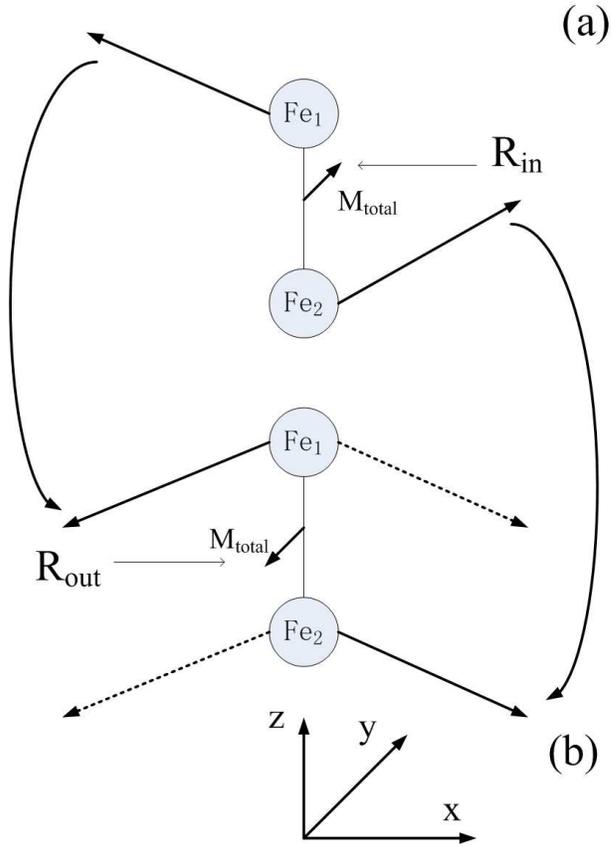}
\caption{Schematic diagram for the coupling between the rotation of
 oxygen octahedra and the resulting magnetization in unit cell in
 BiFeO$_3$. The arrow denote the direction of magnetization.}
\end{figure}

\begin{figure}
\includegraphics{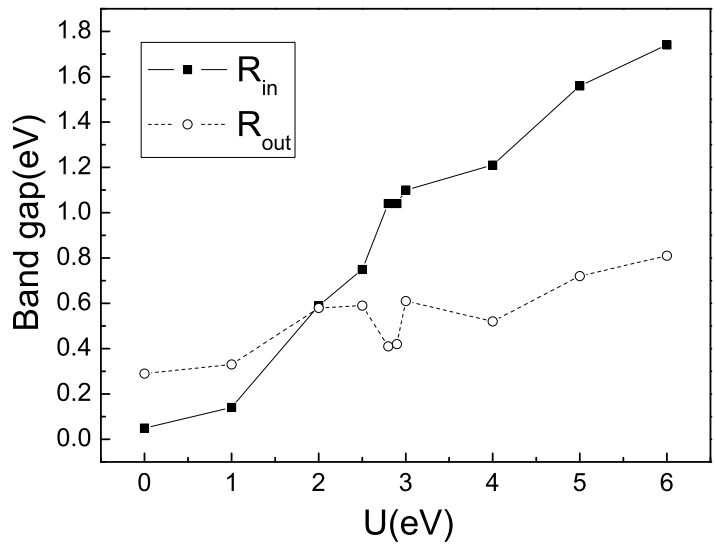}
\caption{Band gap for \textbf{R}$_{in}$ and \textbf{R}$_{out}$ with
respect to $U$.}
\end{figure}

\begin{figure}
\includegraphics{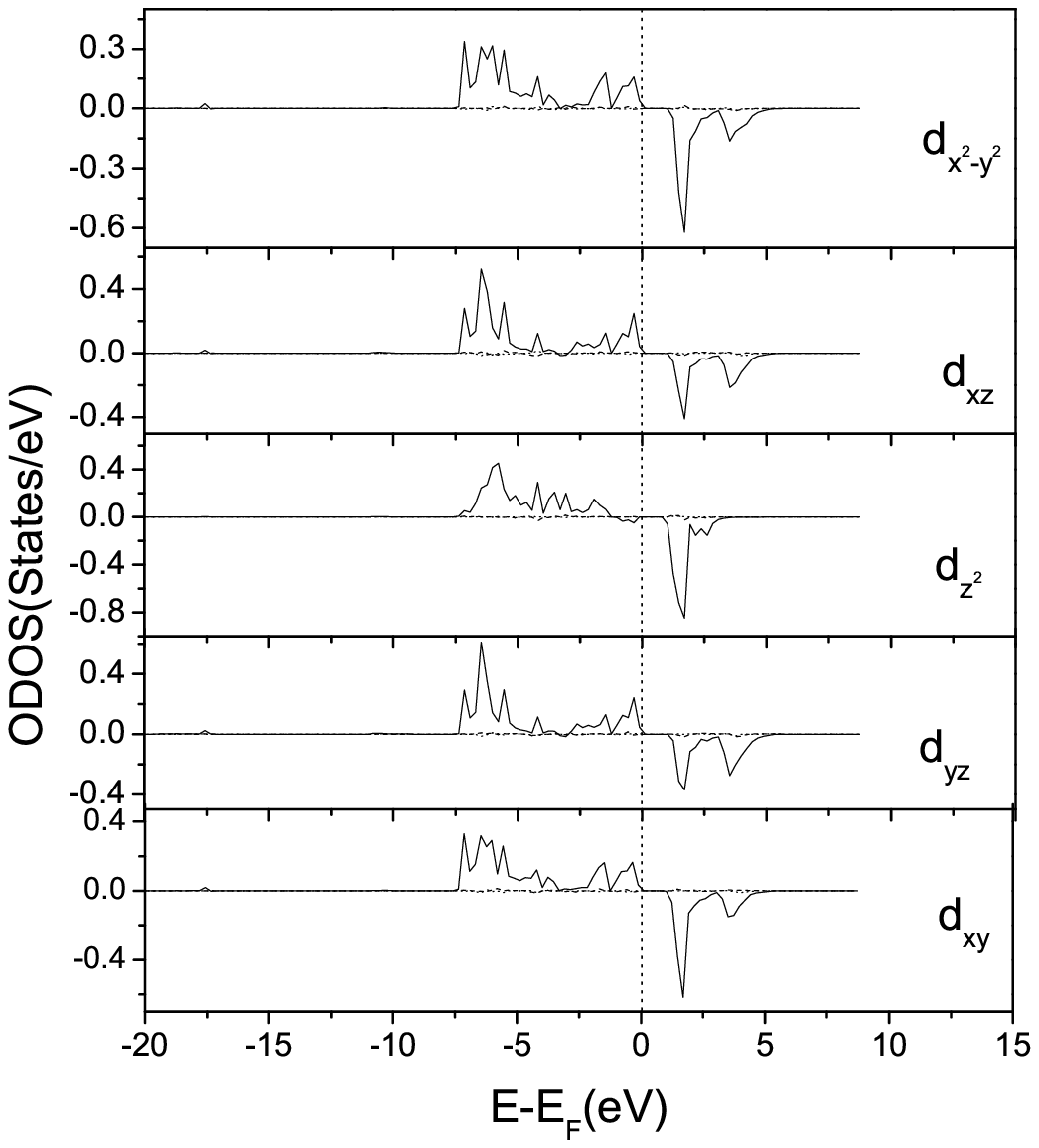}
\caption{ODOS for Fe1 $d_{xy},d_{yz},d_{z^2},d_{xz}$, and
$d_{x^2-y^2}$ orbitals to \textbf{R}$_{in}$.}
\end{figure}

\begin{figure}
\includegraphics{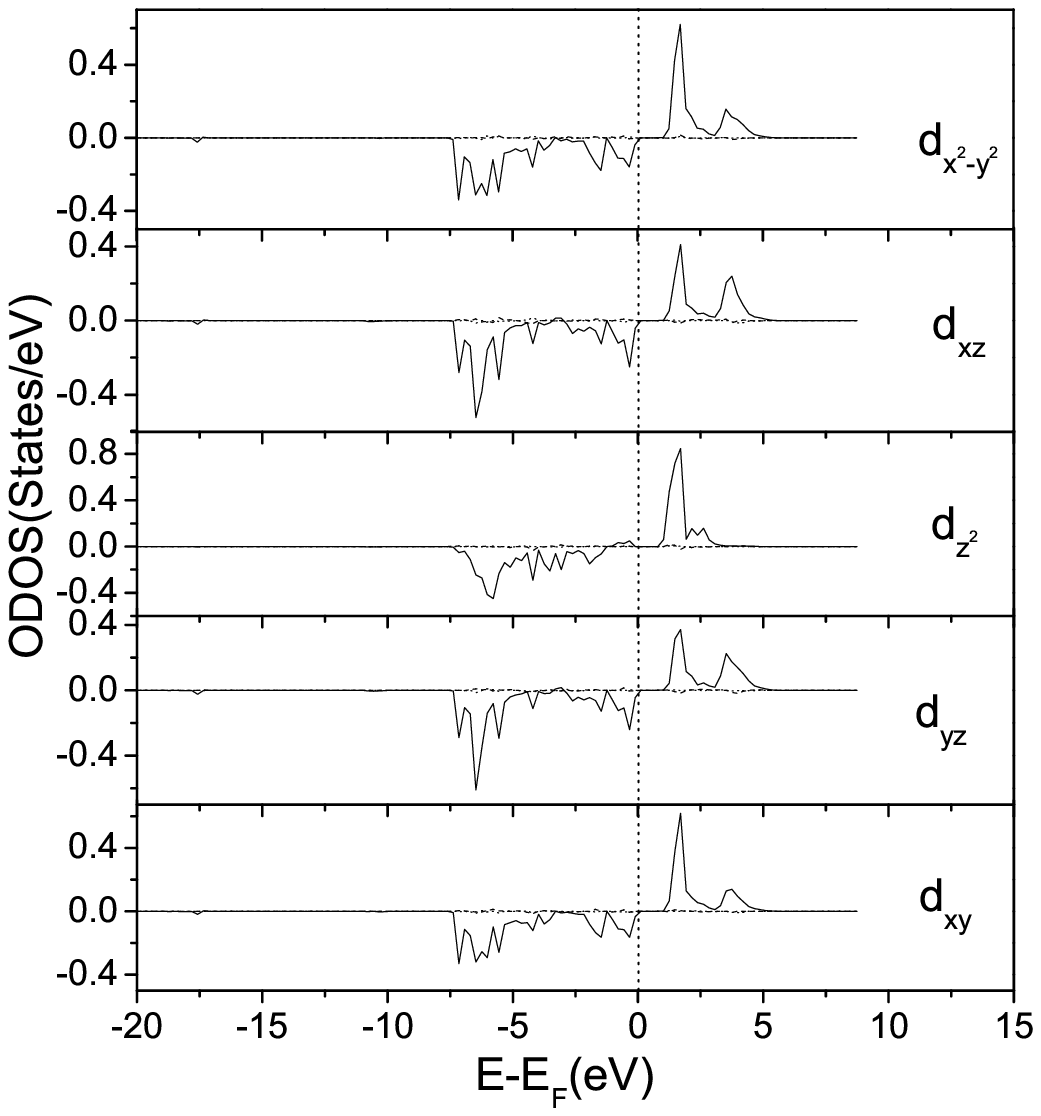}
\caption{ODOS for Fe2 $d_{xy},d_{yz},d_{z^2},d_{xz}$, and
$d_{x^2-y^2}$ orbitals to \textbf{R}$_{in}$.}
\end{figure}

\begin{figure}
\includegraphics{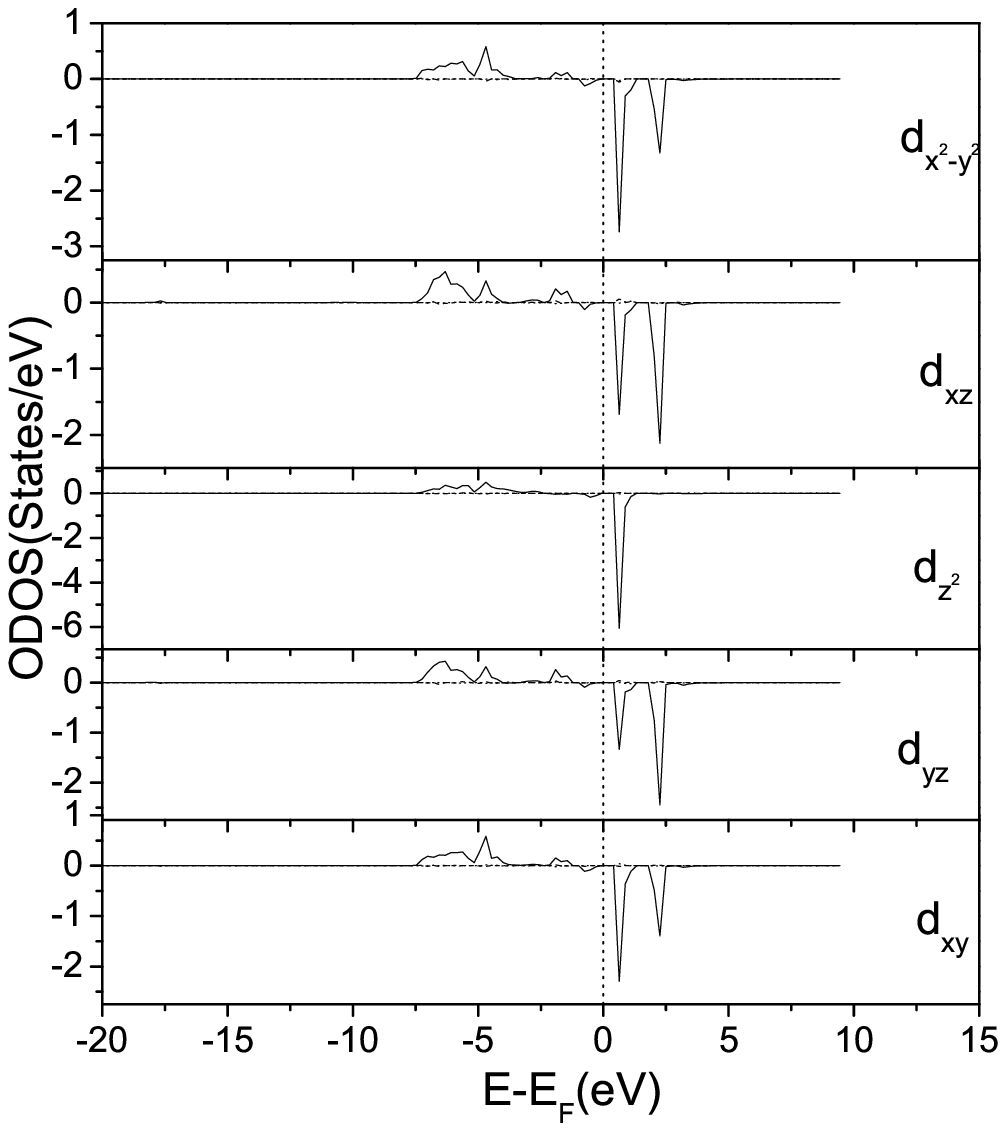}
\caption{ODOS for Fe1 $d_{xy},d_{yz},d_{z^2},d_{xz}$, and
$d_{x^2-y^2}$ orbitals to \textbf{R}$_{out}$.}
\end{figure}

\begin{figure}
\includegraphics{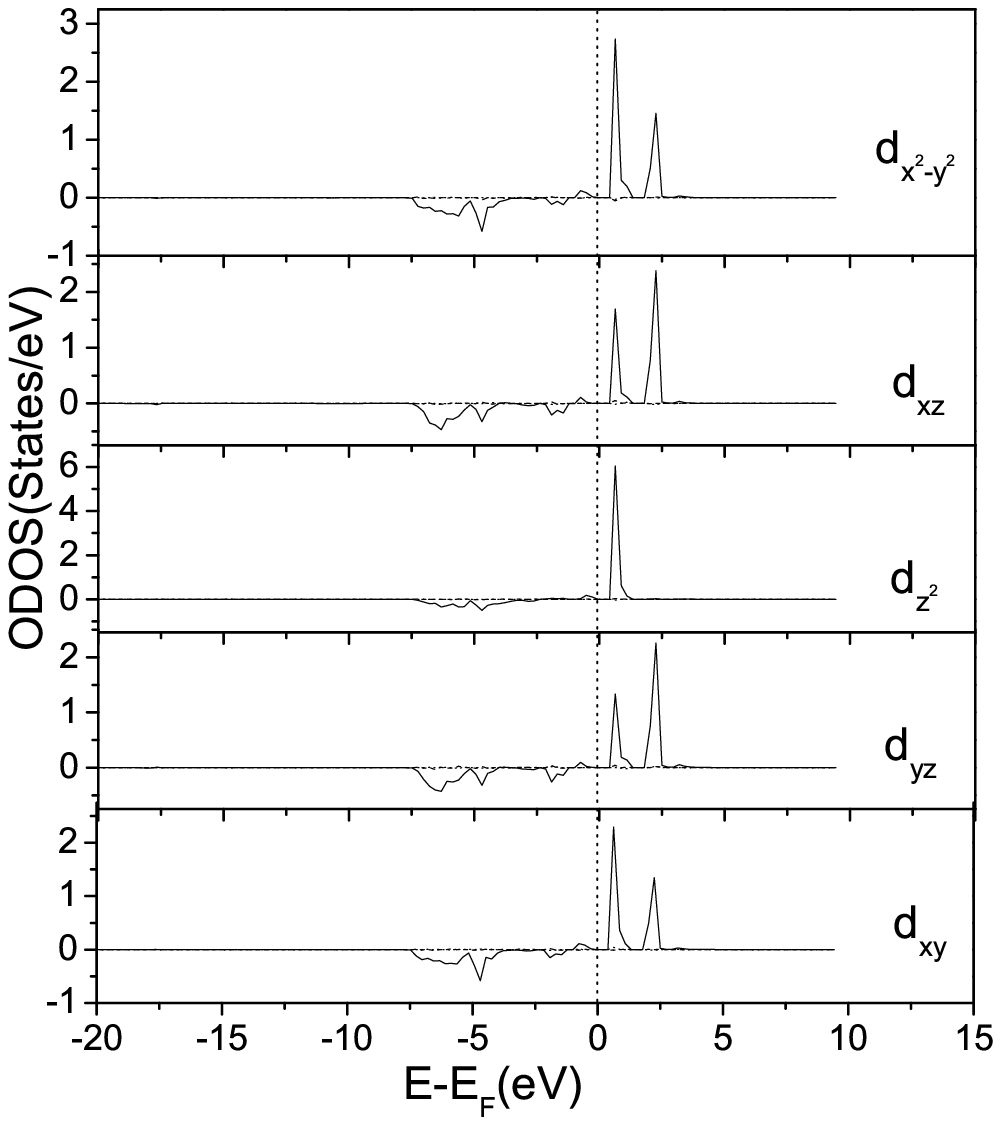}
\caption{ODOS for Fe2 $d_{xy},d_{yz},d_{z^2},d_{xz}$, and
$d_{x^2-y^2}$ orbitals to \textbf{R}$_{out}$.}
\end{figure}

\end{document}